# A Two-Stage Optimization Method for Real-Time Parameterization of PV-Farm Digital Twins

Jong Ha Woo, *Student Member, IEEE,* Qi Xiao, *Student Member, IEEE,* Victor Daldegan Paduani, *Member, IEEE,* and Ning Lu, *Fellow, IEEE*

*Abstract*— Digital twins (DTs) are high-fidelity virtual models of physical systems. This paper details a novel two-stage optimization method for real-time parameterization of photovoltaic digital twins (PVDTs) using field measurements. Initially, the method estimates equivalent irradiance from PV power, voltage, and current data, eliminating the need for direct irradiance sensors. This is crucial for tuning the DT's parameters to actual environmental conditions, thereby improving power prediction accuracy. The second stage focuses on refining these parameters by minimizing discrepancies between measured and predicted outputs. This optimization utilizes the estimated equivalent irradiance as a model input, maintaining synchronization with real-world conditions. Parameter updates are event-trigger, launched when deviations exceed predefined thresholds. This strategy optimizes prediction accuracy and manages communication loads efficiently. Validated with extensive data from a PV farm, this approach outperforms existing methodologies in predictive accuracy and operational efficiency, significantly improving the performance DTs in real-time grid operations.

## I. Introduction

A digital twin (DT) serves as a dynamic, virtual model of a physical system, providing critical insights and enabling advanced predictive capabilities [1]. Recent research has seen a significant surge in the development of power system digital twins (PSDTs) that leverage real-time measurements [2], [3]. This approach aims to replace traditional models that are based on static system snapshots. By using the PSDT as a virtual testbed for emerging technologies, developers, operators, and planners can assess new designs within the current operational context. Additionally, by integrating inputs from forecasting tools, they can proactively predict near-future system operations and swiftly address potential issues during real-time operations. This predictive capability enhances their ability to manage and mitigate risks effectively.

Traditional power system models are typically developed using two main approaches. The first involves constructing standard test systems using standard network topologies and sets of typical parameters [4]-[6]. The second approach captures a snapshot of an existing physical power system to create a static model, effectively preserving a specific moment in time for analysis and simulation. For instance, a Dispatch Training Simulator can accurately replicate a number of operation instances in an actual power grid, initialized from real-time data and historical snapshots [7]-[9].

The fundamental distinction between a conventional power system test bed and a PSDT lies in the dynamic capabilities of the latter. To function as a real-time virtual replica of an actual system, the PSDT must accurately mimic the behavior of the physical system as it changes in real time. Should a significant divergence from the actual system behavior occur, the parameters of the PSDT need to be updated using real-time or near-real-time field measurements. These adjustments should be made without causing sudden disruptions in the simulation outcomes.

To date, while there has been considerable publication on the development of power system test beds and offline model parameterization, research into real-time parameterization of PSDTs has been relatively scarce. Therefore, in this paper, we present a novel two-stage optimization-based method for the real-time parameterization of a photovoltaic digital twin (PVDT). This approach utilizes real-time, high-resolution field measurements as inputs, underscoring the unique requirements of PSDT development and emphasizing the differences from traditional power system test beds. Furthermore, the proposed method avoids the need for irradiance sensors, which are prone to errors due to tilt mismatch and partial cloud shading [10]-[12].

As illustrated in Table I, conventional modeling methods for PV farms are divided into two categories: model-free and model-based approaches. The model-free approach does not incorporate the physical modeling of PV systems; it uses weather data to predict irradiance and PV output power [10]-[12]. This results in a black-box model that does not integrate well with PV control mechanisms and energy management algorithms, making it inadequate for use in digital twin

This material is based upon work supported by the U.S. Department of Energy's Office of Energy Efficiency and Renewable Energy (EERE) under the Solar Energy Technologies Office. Award Number: DE-EE0008770.

Jong Ha Woo, Qi Xiao, and Ning Lu (Corresponding) are with the Electrical & Computer Engineering Department, Future Renewable Energy Delivery and Management (FREEDM) Systems Center, North Carolina State University, Raleigh, NC 27606 USA. (jwoo6@ncsu.edu, qxiao3@ncsu.edu, nlu2@ncsu.edu). Victor Daldegan Paduani is with New York Power Authority (Victor.DaldeganPaduani@nypa.gov).

TABLE I
COMMON MODELING METHODS OF PV FARMS

| | | Model Input | Model Output | Description | Advantage | Disadvantage |
|---|---|---|---|---|---|---|
| Model free based | [10]-[12] | 5 min resolution | Irradiance, PV power output | Estimates variables, including tilt angle, azimuth angles, and albedo, which are related to irradiance. | Parameterize and simulate PV systems without the use of electrical model analysis. | 1) Considers only the power, not the output voltage and current of the PV 2) Performance varies greatly depending on weather conditions. |
| MD model based | [14]-[17] | Manufactural data sheet | PV model parameters | Model the PV system equivalently using the MD model to estimate its parameters. | Accurate electrical modeling of PV becomes possible. | Increased model complexity longer computation times |
| SD model based | [25], [26] (**Benchmark method**) | 5 min resolution | Irradiance | Estimate irradiance by employing an optimization-based approach derived from the equations of the SD model | 1) Information such as tilt angle and albedo are not required. 2) Can consider the uncertainty of irradiance measurement. | Real-time parameterization of the SD model's model parameter was not conducted. |
| | [13], [18]-[23] | Manufactural data sheet | PV model parameters | Estimate model parameter using the SD model's equations and the Lambert W function. However, the uncertainty of the irradiance sensor was not considered. | Can estimate model parameters that correspond to the V-I curve and V-P curve in the manufacturer's datasheet. | 1) Some papers do not consider the time-variability of parameters. 2) If the measured irradiance is inaccurate, the estimated parameters cannot be deemed accurate either. |
| | [24] (**Benchmark method**) | 5 min resolution | PV model parameters | | | |
| | **Proposed method** | 1 sec resolution | Irradiance & PV model parameters | 1) Considers the uncertainty of the irradiance measurement. 2) Estimates model parameters using the SD model's equations and the Lambert W function. | Can maximize the parameterization performance in utility-scale PV data at one-second intervals that are greatly affected by partial shading. | The computation time increases because of co-optimization |

applications.

Nevertheless, to develop a PVDT, it is essential to choose a physics-based PV model capable of accurately predicting the voltage-current (V-I) and voltage-power (V-P) characteristics of PV modules or arrays [13] under varying irradiance levels. This selection enables the PVDT to effectively interface with PV control mechanisms, including real and reactive power regulation and frequency support.

Therefore, we select the model-based approach. Physics-based PV models are typically divided into two types: single-diode (SD) and multi-diode (MD) models. MD models can offer higher prediction precision but also add nonlinearity and complexity to the system [14]-[17]. For instance, the Double Diode model requires seven parameters, and the Triple Diode model requires nine, whereas the SD model only needs five parameters, which will be introduced in Section II.

A PVDT must accurately simulate the behavior of a PV system under various irradiation and temperature conditions. However, employing a more complex model can make real-time parameter adjustments cumbersome. Conversely, using overly simplified models might lead to significant modeling errors, which would require frequent parameter adjustments. Hence there is a trade-off between reducing model complexity, maintaining accuracy, and the necessity for parameter updates. To address this, we have selected the SD model for our PVDT, complemented by an equivalent irradiance calculation. This choice simplifies the problem formulation for parameterization while ensuring a high level of modeling accuracy.

The advantages and disadvantages of the PV modeling approach can also be found in Table I. In [18] and [19], the authors calculate the PV model parameters using the Lambert W function to align the PV model parameters with the voltage-current (V-I) and voltage-power (V-P) curves provided in manufacturer datasheets. The drawback of the approach is that the dynamic variations in PV parameters are not considered [18]-[23]. In [24], the authors use actual operational data from a real PV system and performs parameterization at 5-minute intervals. However, the approach heavily relies on PV measurements, such as irradiance ($G$), temperature ($T$), current of the PV panel ($I_{PV}$), and voltage of the PV panel ($V_{PV}$), without accounting for the uncertainties associated with sensor measurements [13]-[24]. Moreover, sensor data exhibit significant uncertainty; for instance, irradiance sensors may only capture partial information within a PV farm. Relying on sensor measurements alone may result in substantial modeling errors. Note that all those existing models are unable to simulate real-time utility-scale PV farm operations on a second-by-second basis. This limitation stems from the fact that the parameters of the PV models are not fine-tuned in real-time using actual sensor measurements.

Therefore, in this paper, we introduce a two-stage optimization based method for parameterizing a PVDT using real-time measurements and equivalent irradiance calculation. In the first stage, the algorithm estimates equivalent irradiance based on the measured PV power, voltage and current measurements. In this stage, the goal is to produce accurate power prediction using the current PVDT parameters by replacing irradiance sensor inputs with equivalent irradiance. In the second stage, a parameter fine-tuning algorithm will be executed to derive the targeted DT parameters by minimizing the errors between the measurements and the PVDT results using the estimated equivalent irradiance as model inputs.

PVDT parameter updates are based on an event-trigger method. When the predefined thresholds for parameter errors and mismatches between the actual and predicted PV outputs are exceeded, the PVDT parameters will be updated to a new set of values to strike a balance among maintaining high prediction accuracy, reducing overshoots during model parameter transitions, and minimizing communication burden. The effectiveness of our proposed model is validated with field data collected from a PV farm in the North Carolina area, demonstrating its superiority over existing digital twin methodologies.

The primary contributions of this work to the existing body

of literature can be summarized as follows:
- We facilitate real-time parameterization of the PVDT by utilizing high-resolution voltage, current sensor measurements already available at PV farms, enhancing the application of digital twins for real-time operations.
- We selectively updated only the essential parameters required to ensure accurate performance representation of the PVDT, rather than always adjusting all PV parameters in each update.
- We implement event-trigger parameter updates for the model, resulting in significant reductions in computing and communication demands.

To the best of the author's knowledge, there is no comparable research that addresses the challenge of simultaneously optimizing PV irradiance estimation accuracy and five model parameters in real-time. This approach uniquely tackles the uncertainty of irradiance measurements and represents a novel method not previously documented in existing literature.

This paper is structured as follows: Section II elucidates the PV's model parameters and introduces and details the proposed algorithm, contrasting it with existing methodologies. Section III showcases the simulation results, while Section IV concludes the paper, offering final remarks and insights.

## II. METHODOLOGY

In this section, we begin by providing an overview of the parameterization process for PVDT. Following that, we delve into the considerations necessary for PVDT modeling and discuss the methodologies employed to implement parameterization effectively in conducting efficient PVDT.

### A. An Overview of the PVDT Parameterization Process

In this paper, we present a two-stage Particle Swarm Optimization (PSO)-based parameterization method that iteratively calculates the equivalent irradiance ($\tilde{G}$) and PVDT five model parameters (i.e., $R_s, R_{sh}, K_D, I_{ph0}, I_s$) for accurately tracking and predicting the current ($I_{PV}$) and voltage ($V_{PV}$) of the PV system in real-time operation.

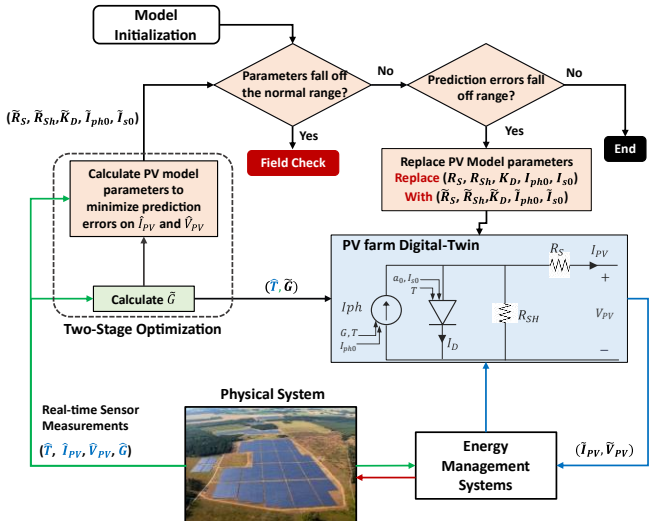

**Fig. 1.** An overview of the PVDT parameterization process.

As shown in Fig. 1, a real-time PVDT provides a virtual platform for the Energy Management System (EMS) to predict future outputs and assess various control strategies. In PV farms, irradiance sensors are sparsely deployed, making it impractical to directly use the irradiance sensor measurement ($\hat{G}$) for real-time power output estimation of the PV farm. To address this, we propose calculating the equivalent irradiance ($\tilde{G}$) using real-time voltage ($\hat{V}_{PV}$), current ($\hat{I}_{PV}$), and temperature ($\hat{T}$) measurement so that $\tilde{G}$ instead of $\hat{G}$ can be used as the input to the PVDT mode.

Operational conditions in a PV farm vary due to factors such as changes in solar angle, adjustments to the tracking mechanism, and modifications implemented by controllers. Therefore, relying on a single set of PVDT parameters to model all operational conditions can result in significant modeling errors under certain circumstances. Thus, when discrepancies between the predicted current ($\tilde{I}_{PV}$) and voltage ($\tilde{V}_{PV}$) and the measured values (i.e., $\hat{V}_{PV}$ and $\hat{I}_{PV}$) exceed predetermined thresholds, it becomes necessary to recalculate the parameters of the PVDT model.

### B. PVDT Modeling Considerations

As shown in Fig. 1, a lumped SD based PV model is used to model PV panels for its simplicity. The model consists of five key parameters: the diode ideality coefficient ($K_D$), the diode saturation current coefficient ($I_{s0}$), the series resistance ($R_s$), the shunt impedance ($R_{sh}$), and the photocurrent source coefficient ($I_{ph0}$). As introduced in [25]-[28], the PV model can be described as follows.

$$\hat{I}_{PV} = I_{ph} - I_D - I_{R_{sh}} \tag{1}$$

$$\hat{I}_{PV} = I_{ph} - I_s \left(e^{\frac{\hat{V}_{PV} + \hat{I}_{PV} R_s}{a}} - 1\right) - \frac{\hat{V}_{PV} + \hat{I}_{PV} R_s}{R_{sh}} \tag{2}$$

$$I_{ph} = \tilde{G} I_{ph0}\left(1 + (\hat{T} - 25)\alpha_{Isc}\right) \tag{3}$$

$$a = K_D \frac{kT_{STC}}{q} N_s \left(\frac{\hat{T} + T_{FP}}{T_{STC}}\right) \tag{4}$$

$$I_s = I_{s0} \left(\frac{\hat{T} + T_{FP}}{T_{STC}}\right)^3 e^{47.1\left(1 - \frac{T_{STC}}{\hat{T} + T_{FP}}\right)} \tag{5}$$

$$\omega = W\{I_{ph} \frac{e}{I_s}\} \tag{6}$$

$$\tilde{V}_{PV} = \left(1 + \frac{R_s}{R_{Sh}}\right) a (\omega - 1) - R_s I_{ph} \left(1 - \frac{1}{\omega}\right) \tag{7}$$

$$\tilde{I}_{PV} = I_{ph}\left(1 - \frac{1}{\omega}\right) - \frac{a(\omega - 1)}{R_{Sh}} \tag{8}$$

$$\tilde{P}_{PV} = \tilde{V}_{PV} \times \tilde{I}_{PV} \tag{9}$$

where $\hat{T}$ is the Celsius temperature at the PV array, $T_{STC}$ denotes the reference temperature at standard test condition (STC) (i.e., $T_{STC} = 298.15$ K), and $T_{FP}$ is the temperature at freezing point (FP) (i.e., $T_{FP} = 273.15$ K); $I_D$ is the current flowing through the diode, $I_{R_{sh}}$ is the current flowing through the shunt resistance, and $I_s$ is the diode saturation current; $\alpha_{ISC}$ is the temperature coefficients of the short-circuit current; $k$ is the Boltzmann constant and $q$ is the electron charge; $N_s$ is the series connected cells in the PV unit; $\omega$ is an auxiliary parameter related to $I_{ph}$ and $I_s$ via the Lambert W function; $\tilde{P}_{PV}$ is the maximum power point (MPP); $W$ denotes the Lambert $W$ function for estimating the PV voltage and current at the MPP, $\tilde{V}_{PV}$ and $\tilde{I}_{PV}$, respectively.

Thus, using (1)-(9), assuming that the five PV model parameter (i.e., $R_s, R_{sh}, K_D, I_{ph0}, I_s$) are known, the PVDT can predict the PV farm output in the MPP tracking (MPPT) mode using $\tilde{G}$ and $\hat{T}$ as inputs.

### C. Irradiance Estimation

When a PV farm is not used for providing grid services, utilizing average irradiance sensor inputs to predict low resolution (i.e., 15-minute or hourly) PV outputs would have been sufficient [24]. However, in a high-renewable penetration grid, it is anticipated that grid-scale PV farms will be required to offer high-quality ancillary services, such as providing regulation or fast frequency responses [29], at a level comparable to conventional generators. Thus, it is crucial for the PVDT to predict the PV output at a second-by-second resolution to predict the power reference accurately.

As shown in Fig. 2, the correlation between the 1-second measured PV farm output ($\hat{P}_{PV}$) and the irradiance sensor measurement ($\hat{G}$) is significantly lower when compared to the 15-minute case. This is because there are only a limited number of sensors installed in a PV farm. These sensors provide accurate measurements only for nearby PV panels. However, in the case of utility-sized PV farms, which encompass tens of thousands of panels over extensive areas, these irradiance measurements fall short in accounting for factors such as partial shading and suboptimal tilt angle, accumulation of dirt, and dynamic adjustments in panel tilt angles.

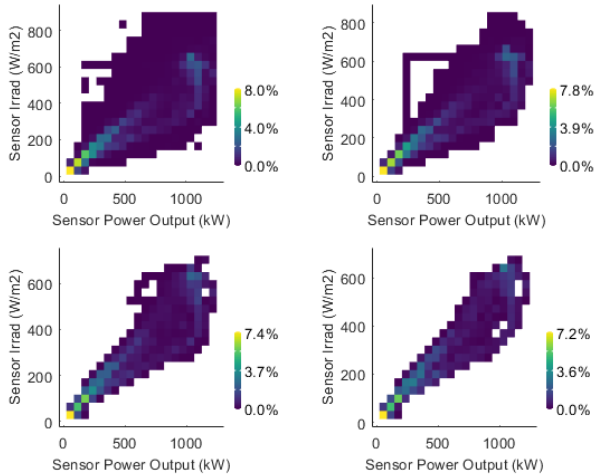

**Fig. 2.** Correlation between measured PV output and Irradiance. Data resolution: (Top-Left) 1-second, (Top-Right) 1-minute, (Bottom-Left) 5-minutes, (Bottom-Right) 15-minutes.

Thus, the irradiance sensor measurements are not suitable to be directly used by the PVDT for accurately predicting high-resolution (1-minute or less) PV farm voltage, current, and power outputs. In [25] and [26], assuming that the five parameters of the SD-based PV model are known, the authors propose to estimate an equivalent irradiance ($\tilde{G}$) by minimizing $|f_1(\tilde{G})|$.

By substituting (3), (4), and (5) into (2) using Kirchhoff's Current Law (KCL), we have

$$f_1(\tilde{G}) = -\hat{V}_{PV} - \hat{I}_{PV}R_s +$$
$$a \ln \left[ \frac{\tilde{G}I_{ph0}[1 + \alpha_{Isc}T_{STC}\left(\left(\frac{\hat{T}+T_{FP}}{T_{STC}}\right)-1\right)] - \hat{I}_{PV} - \left(\hat{V}_{PV} + \frac{\hat{I}_{PV}R_s}{R_{sh}}\right)}{I_{s0}\left(\frac{\hat{T}+T_{FP}}{T_{STC}}\right)^3 e^{47.1\left(1 - \frac{T_{STC}}{\hat{T}+T_{FP}}\right)}} \right] \quad (10)$$

Note that if the solar irradiance ($G$) is known, we can also use (10) to estimate the PVDT model parameters, namely $K_D, I_{s0}, R_s, R_{sh}$, and $I_{ph0}$. However, if we opt to directly use irradiance sensor measurements ($\hat{G}$) for PVDT parameter estimation, the accuracy of these parameters ($R_s, R_{sh}, K_D, I_{ph0}, I_{s0}$) may be compromised due to the low correlation between $\hat{G}$ and PV power output, as shown in Fig. 2. To address this problem, we introduce a two-stage optimization method in the following section to iteratively optimize the equivalent irradiance and the five PVDT model parameters, offering a robust solution for PVDT parameterization using real-time measurement.

### D. Two-stage PSO

The problem presented in this paper cannot be addressed simply by solving equations, as it does not follow the typical form of $f(x) = 0$. The key challenge lies in the inherent inaccuracies of measurements, not only for irradiance, as emphasized in the manuscript, but also for current and voltage. Real-world data is often rounded to a few decimal places, leading to errors, and standard power system SCADA data typically has 1-2% error in current and voltage measurements. Consequently, obtaining an exact analytical solution for $f(x) = 0$ is infeasible, necessitating the optimization.

While both classical (e.g., Linear Programming, Interior Point Method) and heuristic methods (e.g., Particle Swarm Optimization (PSO), Ant Colony Optimization) were considered, PSO was selected for following reasons: (1) Since the objective function is nonlinear, Linear Programming (LP) is not a viable option.; (2) Nonlinear methods such as Interior Point are computationally intensive due to the need for differentiation and are prone to getting stuck in local minima depending on the initial values; and (3) Although more sophisticated methods like Ant Colony Optimization could be applied, the problem does not require such complexity. Our goal was to choose the simplest, fastest, and most robust algorithm for real-time parameterization, rather than focusing on the algorithm itself.

In PSO [30], the Personal Best ($P_{best}$) is the best solution found by an individual particle and the Global Best ($G_{best}$) is the best solution found by all particles in the swarm. Using $\hat{I}_{PV}$, $\hat{V}_{PV}, \hat{T}$ data as input, we propose a two-stage PSO method for parameterizing a PVDT in real-time by iteratively estimating $X_1$, representing the equivalent irradiance ($\tilde{G}$) and $X_2$, representing the five PV SD model parameters. Furthermore, measurements transmitted via SCADA ensure accurate and timely data for the optimization process. The proposed model is illustrated in Fig. 3.

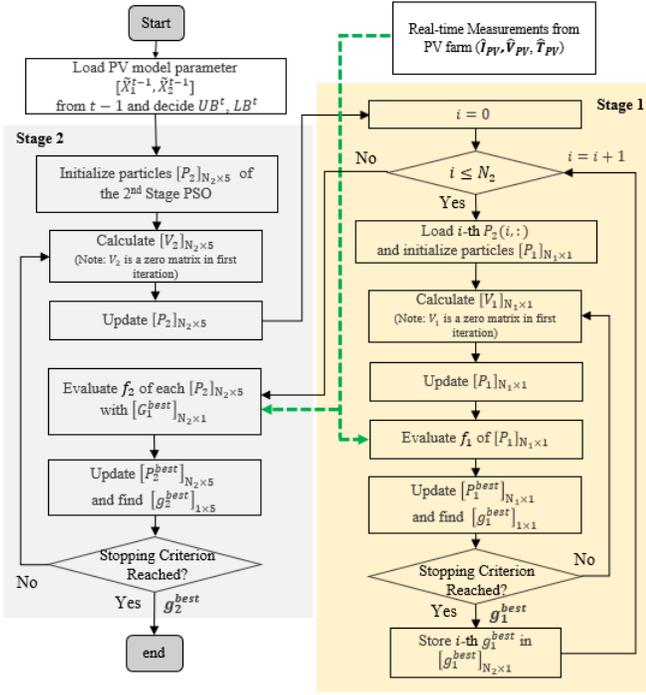

**Fig. 3.** Two-stages PSO Algorithm to find model parameters.

To initialize the algorithm, generate $N_1$ random particles $P_1$ for $X_1$, where $X_1 \subseteq [LB_1, UB_1]$ and generate $N_2$ random particles $P_2$ for $X_2$, where $X_2 \subseteq [LB_2, UB_2]$. Note that MATLAB offers model parameters for 21,187 photovoltaic (PV) modules. Using those data as inputs, the parameter ranges (i.e., $[LB_2, UB_2]$) for the five PV model parameters can be obtained, as shown in Fig. 4. The irradiance typically ranges from 0 to 1,000 $W/m^2$, so $X_1$ will generate random values between [0, 1000], and $X_2$'s initial values will be generated between the min and max values shown in Fig. 4.

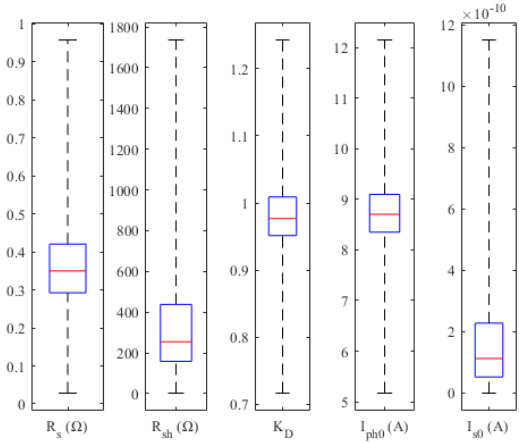

**Fig. 4.** Feasible range of five model parameters derived from a dataset of 21,187 PV modules.

As illustrated in Fig. 3, the first stage estimates $\tilde{G}$ assuming ($R_s, R_{sh}, K_D, I_{ph0}, I_{s0}$) remain unchanged. Thus, for each set of PV model parameters in $X_2$, PSO is used for solving (10) by

$$\min_{\tilde{G}} |f_1(X_1)| \qquad (11)$$

Although $\tilde{G}$ is calculated to satisfy the KCL, the resulting $\tilde{V}_{PV}$ and $\tilde{I}_{PV}$ may differ significantly from the measured voltage and current of the PV farm, especially when the control logic inside the PV controller varies. Thus, in the second stage PSO, $\tilde{G}$ is used as the input to estimate ($R_s, R_{sh}, K_D, I_{ph0}, I_{s0}$) by minimizing the errors between calculated PV voltage and current ($\tilde{V}_{PV}$ and $\tilde{I}_{PV}$) and the measured ones ($\hat{V}_{PV}$ and $\hat{I}_{PV}$), assuming that the PV system is operating in MPPT mode.

The second stage objective function is defined as

$$\min_{R_s, R_{sh}, K_D, I_{ph0}, I_{s0}} f_2(X_2)$$

$$\begin{aligned} f_2(X_2) &= \left|\hat{I}_{PV} - \tilde{I}_{PV}\right| + \left|\hat{V}_{PV} - \tilde{V}_{PV}\right| \\ &= \left|\hat{I}_{PV} - I_{ph}\left(1 - \frac{1}{\omega}\right) - \frac{a(\omega-1)}{R_{Sh}}\right| \\ &+ \left|\hat{V}_{PV} - \left(1 + \frac{R_S}{R_{Sh}}\right)a\,(\omega-1) - R_S I_{ph}\left(1 - \frac{1}{\omega}\right)\right| \end{aligned} \qquad (12)$$

In stage 2, each particle, $X_2$, is evaluated using (12). The dual-stage optimization process can then identify a pair of [X1, X2] that meet the KCL law in (11) and minimize the errors between the predicted $\tilde{V}_{PV}$ and $\tilde{I}_{PV}$ and the actual measured values.

In the PSO algorithm, the velocity vector is represented by acceleration coefficients ($c_1, c_2$) and inertia weight ($w$), and the function for determining velocity ($V_i^k$) and position ($X_i^k$) can be found in equation (13).

$$V_i^{k+1} = wV_i^k + c_1 r_1 (P_{best\,i}^k - X_i^k) + c_2 r_1 (G_{best\,i}^k - X_i^k) \qquad (13)$$

In this experiment, the PSO coefficients were set to $c_1 = 0.4$, $c_2 = 0.4$, and $w = 0.5$, with r1 and r2 being random numbers between 0 and 1.

*E. PVDT Parameter Selection and Adjustment*

As shown in Fig. 5, the V-I and V-P curves given by the manufacturers are obtained by conducting experiments under standard weather conditions for a single PV module rather than for a large utility-scale PV farm. Consequently, when applying these curves to derive model parameters for a large PV farm comprising thousands of modules, time-varying effects such as partial shading and tracking, cannot be fully accounted for by a single set of model parameters. Therefore, adjustments to the PVDT parameters are necessary to ensure accurate predictions.

When fine-tuning the PVDT model parameters in real-time, our objectives are two-fold: reduce the frequency of parameter adjustments and minimize the number of parameters that need to be adjusted. To minimize the need for frequent parameter updates, it is critical to begin with an optimal set of ($R_s, R_{sh}, K_D, I_{ph0}, I_{s0}$) that can produce satisfactory predictions under most operation scenarios. Thus, for $N$ data points on the V-I curves and V-P curves given in the PV module datasheet [18]-[23], we use the RMSE minimization method introduced in [18] to calculate an optimal set of PVDT parameters as

$$\min_{R_s, R_{sh}, K_D, I_{ph0}, I_{s0}} \text{RMSE} = \sqrt{\frac{1}{N}\sum_{i=1}^{N}\left(\tilde{I}^{(i)}{}_{PV} - \hat{I}^{(i)}{}_{PV}\right)^2} \qquad (14)$$

where $\tilde{I}^{(i)}{}_{PV}$ and $\hat{I}^{(i)}{}_{PV}$ are the estimated and the measured PV

module current at point $i$, respectively.

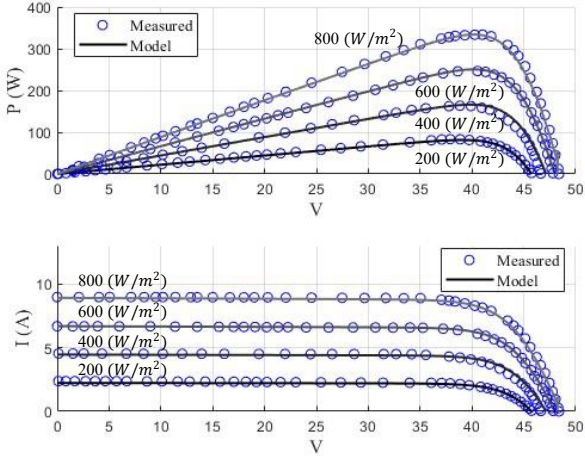

**Fig. 5.** The characteristics of PV module: V-P Curve (up) and V-I Curve (down).

In this paper, the optimal set of parameters we obtained are as follows: $X_2^{opt} = [R_S, R_{Sh}, K_D, I_{ph0}, I_{s0}] = [0.279, 216.990, 1.086, 11.134, 3.405 \times 10^{-10}]$. The RMSE between the black line and the blue measurement values on the V-I Curves (see Fig. 5) is 0.149 A.

To reduce the number of parameters requiring adjustment, we assess the parameter sensitivity. Given that $I_{ph0}$ is a panel constant and $I_{s0}$ and $K_D$ are diode constants, we prioritize adjusting the values of $R_s$ and $R_{sh}$ first. We consider modifying the constants $I_{ph0}, I_{s0}$, and $K_D$ only if significant discrepancies remain after updating $R_s$ and $R_{sh}$.

*F. PV Model Parameter Update Strategies*

As a simplified representation, the SD model cannot depend on just one set of model parameters to accurately predict PV farm outputs under various operational conditions. Therefore, it is critical to identify when to update the PVDT model parameters to ensure prediction errors remain within acceptable limits, while also minimizing communication requirements and computational costs.

In this paper, we apply two parameter update strategies: *fixed-time-interval* updates and *threshold-trigger* updates, as illustrated in Fig. 6. The fixed-time-interval approach schedules updates at set intervals, for instance, every minute or every 30 minutes. The threshold-trigger (event-trigger) approach initiates parameter updates only when the error between estimated ($\tilde{V}_{PV}, \tilde{I}_{PV}$) and measured ($\hat{V}_{PV}, \hat{I}_{PV}$) is above an established threshold. A comparison of the two strategies is presented in Section III.

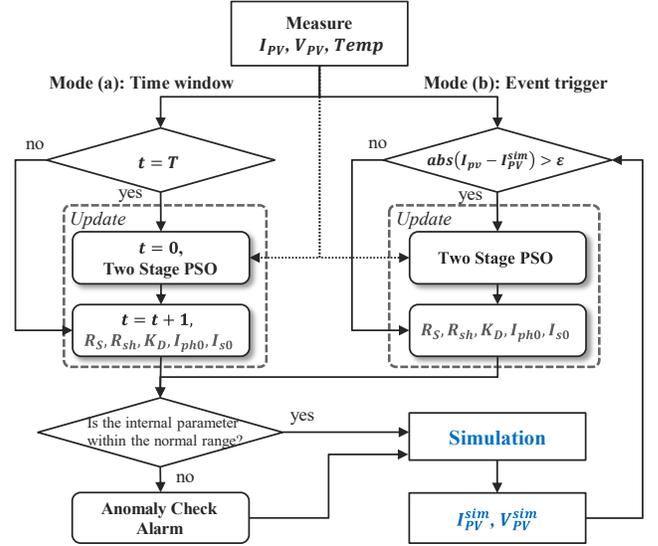

**Fig. 6.** Update criteria: (a) Fixed-time-interval, and (b) Event-trigger.

## III. SIMULATION RESULTS

To validate the performance of the proposed 2-stage PSO based PVDT parameterization algorithm, we use the 1-second dataset collected in November 2022 from a utility-scale PV farm operated by Strata Clean Energy at North Carolina area. The PV farm runs in MPPT modes in normal operation and will execute curtailment when its interconnection agreement is exceeded; however, scenarios involving curtailment are excluded from our simulation case. The corresponding PVDT model of the PV farm runs in the Simulink environment and is controlled by the Perturb & Observe algorithm when in MPPT mode. Additionally, while the dataset had 1 second resolution, the optimization was performed at a 10-second resolution.

*A. Performance Comparison of the Proposed Method*

In this section, we evaluate the performance of our proposed two-stage PSO-based parameterization methodology against three benchmark methods. These include: fixed PVDT parameters with measured irradiance as inputs (Base method), fixed PVDT parameters with estimated equivalent irradiance as inputs (Method 1), and real-time estimated PVDT parameters with measured irradiance as inputs (Method 2). Note that the fixed PVDT parameters were obtained as described in in Section II. E.

Fig. 7 shows the distribution of the PVDT model parameters, based on data from a three-hour period on a specific day in October 2022. Table II provides a summary of the performance

TABLE II
COMPARISON OF RESULTS ACROSS THREE OPTIMIZATION METHODS

| | Real-time Parameterization of the SD model | Irradiance (measured or estimated) | Mean Absolute Percentage Error (MAPE) | | | Root Mean Square Error (RMSE) | | |
|---|---|---|---|---|---|---|---|---|
| | | | Δ I (%) | Δ V (%) | Δ P (%) | Δ I (A) | Δ V (V) | Δ P (kW) |
| Base Case | $X_2^{opt}$ calculated by (13) | measured | 37.36 | 1.88 | 38.39 | 333.41 | 24.67 | 354.58 |
| Method 1 | | estimated | **1.05** | 0.92 | **0.14** | 9.15 | 11.41 | **1.38** |
| Method 2 | Real-time estimated by (12) | measured | 2.13 | **0.42** | 1.81 | 64.75 | **6.04** | 66.52 |
| Proposed | Co-optimized by 2-stage PSO | | 0.25 | 0.24 | 0.06 | 2.93 | 2.95 | 0.59 |

comparison results.

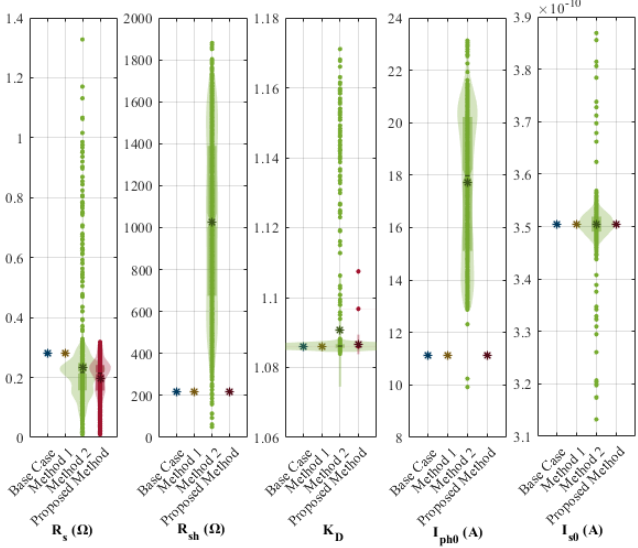

**Fig. 7.** Five model parameters range according to each method.

Fig. 8 presents an example of the time-series plot of the current, voltage, and power outputs of the PVDT, also based on the same data. From the simulation results, we make the following observations:

- **Base Case:** As shown in Table II and Fig. 8, the simulation results of the Base case is the worst among the four methods, showing very large discrepancies when compared to the field current, voltage and power measurements.
- **Method 1**: Method 1 uses the same PVDT model parameters as the Base case. The green lines in Figs. 8(a) and 8(c) show that substituting actual irradiance measurement with estimated equivalent irradiance significantly improves the accuracy of current and power estimation. However, while the voltage estimation error in Method 1 is smaller compared to the Base case, it is considerably exceeding the errors observed in Method 2 and the proposed method, as shown in Table II.
- **Method 2:** By estimating the PVDT parameters in real-time, Method 2 outperforms Method 1 in voltage estimation, as shown in Fig. 8(b). However, its current estimation shows very large discrepancies, as shown in Fig. 8(a). Additionally, both its current and voltage estimations show high spikes, indicating that significant deviations between actual and measured irradiance affect the efficacy of solely adjusting PV model parameters for predicting PV current and voltage measurements accurately. Thus, when using Method 2, the calculated $R_s, R_{sh}$ and $I_{ph0}$ can exceed the parameter range showed in Fig. 4.
- **Summary of Methods 1&2:** Those results show that even if parameters satisfying KCL are found in Method 1 or Method 2, the $\tilde{V}_{PV}$ and $\tilde{I}_{PV}$ estimated by the PVDT may differ from $\hat{V}_{PV}$ and $\hat{I}_{PV}$. Therefore, it is essential to concurrently calculate equivalent irradiance and optimize the SD PV model parameters in real-time, as the proposed model does.

- **The proposed 2-stage PSO parametrization method.** By co-optimizing model parameters and calculating equivalent irradiance in real time, the proposed method shows the best performance, as shown in Table II. Additionally, the PV model parameters obtained are within a smaller range, compared to Method 2, as shown in Fig. 7. Note that letting PVDR parameters vary drastically in real-time simulation is undesirable. In [24], the authors recommend that $R_s$ can fluctuate between about 0.5 and 9, $R_{sh}$ between 500 and 4000 and $I_{ph0}$ between 0.8 and 5. Therefore, the proposed parameterization methodology outperformed the other three benchmark methods.

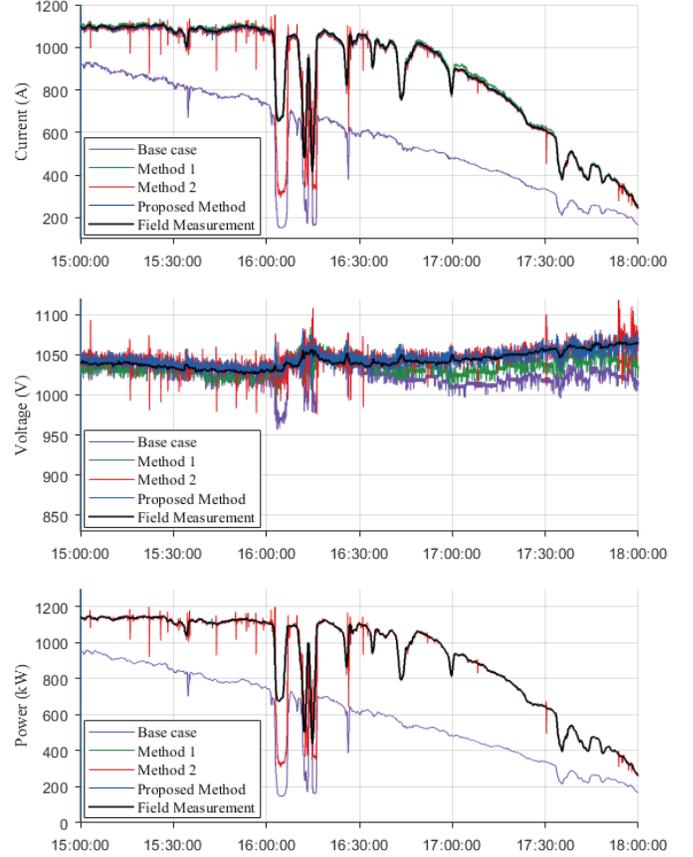

**Fig. 8.** Simulink simulation results according to each method

In Fig. 8, the red line (representing Method 2) exhibits significant spikes, which are attributed to the transient response caused by rapid changes in parameters in real-time. In this simulation, the interval at which parameters are updated is 10 seconds, during which the system experiences a transient state before reaching a steady state. This paper defines the overshoot and undershoot occurring during the transient state as Transient Performance Indices (TRI), as described in Equation (14).

$$\text{TRI} = \max\left(\frac{MaxV - SSV}{SSV}, \frac{SSV - MinV}{SSV}\right) \times 100\% \qquad (14)$$

where SSV is steady-state value (SSV), MaxV and MinV indicate maximum and minimum value during the transient state, respectively.

The distribution of TRI throughout the simulation time is depicted in Fig. 9, indicating that higher TRI values correlate with more frequent occurrences of significant overshoot or undershoot within the system. The overshoots in the base case and Method 1, where parameters are fixed regardless of time, are less than in the other two methodologies (Method 2 and Proposed Method) where parameters change over time. However, the proposed method, which has smaller real-time parameter fluctuations, exhibited less overshoot compared to Method 2.

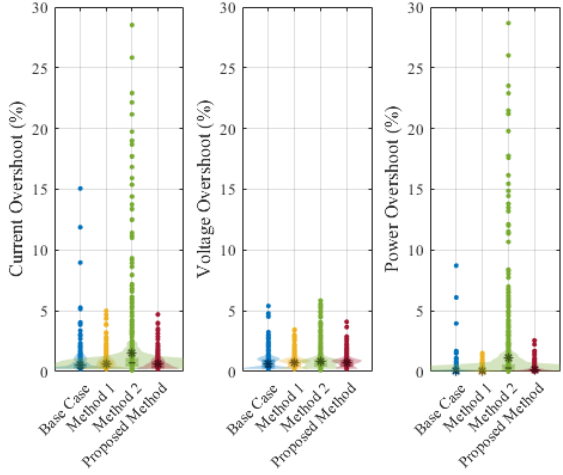

**Fig. 9.** TRI manifestations in simulations according to each method

### B. Comparing PVDT Parameter Update Strategies

This section evaluates the proposed algorithm's performance across two distinct parameter update strategies: fixed parameter update intervals and event-trigger updates. Table III presents the outcomes for five varied fixed update intervals alongside one event-trigger scenario. From the results, we have the following observations:

- Among the five fixed parameter cases, the 10-second update case yields the best results as expected. However, this requires frequent communication between the PVDT and the actual environment, leading to higher communication and computing costs. Moreover, such frequent updates are largely unnecessary because the model parameters change only slightly within a 10-second timeframe.
- If the parameters are updated every 60 minutes, the modeling errors for current and voltage increase by 0.65 and 0.71, respectively, while the power estimation error only rises by 0.1. This indicates that for applications where the primary focus of the PVDT is on power output, updating the model at a slower pace is deemed acceptable.
- The Event-trigger method only updates the PVDT model parameters when significant errors in current and voltage modeling are detected, as shown in Fig. 6. In our experiment, we set the error threshold at 0.5%. As shown in Table III, using the event-trigger strategy resulted in 58 parameter updates over a three-hour period and achieved comparable performance with that achieved with 1-minute updates, yet it requires only one-sixth the number of parameter updates.
- The results of the five model parameters for the 10-second, 1-minute, and event-trigger cases are plotted in Fig. 10. The parameter updated frequently is $R_S$, showing its sensitivity on PVDT modeling performance is much higher than the remaining four parameters. In some cases, $R_{Sh}$ also requires an update, mainly when PVDT performance cannot be guaranteed by adjusting $R_S$ alone, necessitating incremental adjustments to the $R_{Sh}$ parameter. Additionally, the panel constants $K_D, I_{ph0}$, and $I_{s0}$ estimated from the manufacturer's datasheet to maintain the $X_2^{opt}$ hourly, will undergo minor changes when adjustments to physical parameters alone are insufficient to ensure PVDT performance. Especially, note that in the case of $I_{ph0}$ and $I_{s0}$, no changes were observed and the values $[11.134, 3.405 \times 10^{-10}]$ corresponding to $X_2^{opt}$ were maintained; therefore, the remaining two parameters ($I_{ph0}$ and $I_{s0}$) are omitted in Fig. 10.

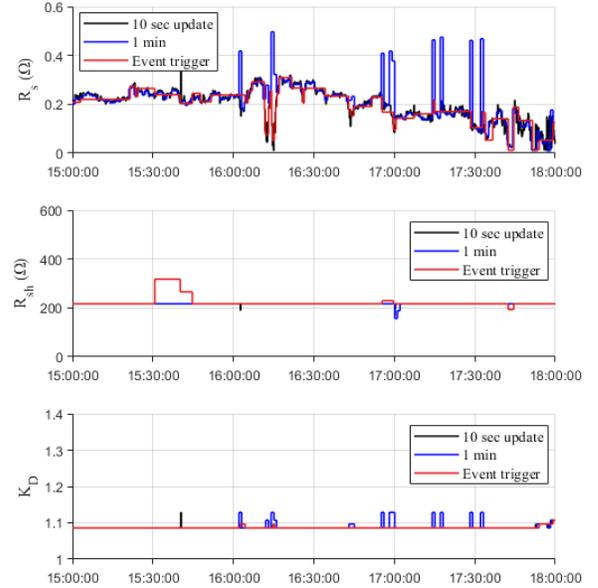

**Fig. 10.** Changes in Five model parameters according to update criteria.
\* No changes were observed in $I_{ph0}$ and $I_{s0}$, so they are omitted.

**TABLE III**
**COMPARISON OF RESULTS ACROSS UPDATE STRATEGIES**

| Update Criteria | Update Counts In 3 hours (10,800 sec) | Δ I (absolute percentage error) | | | Δ V (absolute percentage error) | | | Δ P (absolute percentage error) | | |
|---|---|---|---|---|---|---|---|---|---|---|
| | | Mean (%) | Min (%) | Max (%) | Mean (%) | Min (%) | Max (%) | Mean (%) | Min (%) | Max (%) |
| **10 sec** | **1,079** | **0.25** | **5.33E-04** | **1.57** | **0.24** | **3.84E-04** | **0.75** | **0.06** | **1.53E-05** | **1.43** |
| **1 min** | **359** | **0.33** | **1.02E-04** | **1.74** | **0.34** | **3.80E-04** | **1.75** | **0.06** | **1.90E-04** | **0.85** |
| 5 min | 71 | 0.55 | 4.13E-04 | 2.51 | 0.58 | 2.09E-03 | 2.15 | 0.07 | 7.22E-05 | 1.15 |
| 10 min | 35 | 0.58 | 1.71E-03 | 3.21 | 0.61 | 1.84E-03 | 2.57 | 0.08 | 1.21E-04 | 1.29 |
| 15 min | 23 | 0.59 | 2.18E-03 | 2.67 | 0.62 | 1.25E-03 | 2.23 | 0.08 | 2.40E-05 | 1.29 |
| 30 min | 11 | 0.61 | 1.46E-03 | 3.51 | 0.65 | 4.44E-04 | 2.60 | 0.09 | 8.03E-05 | 0.82 |
| 60 min | 2 | 0.65 | 4.84E-05 | 2.87 | 0.71 | 3.44E-03 | 2.27 | 0.10 | 1.34E-04 | 0.80 |
| **Event trigger** | **58** | **0.36** | **2.05E-03** | **1.38** | **0.36** | **1.57E-04** | **1.30** | **0.07** | **1.42E-04** | **0.93** |

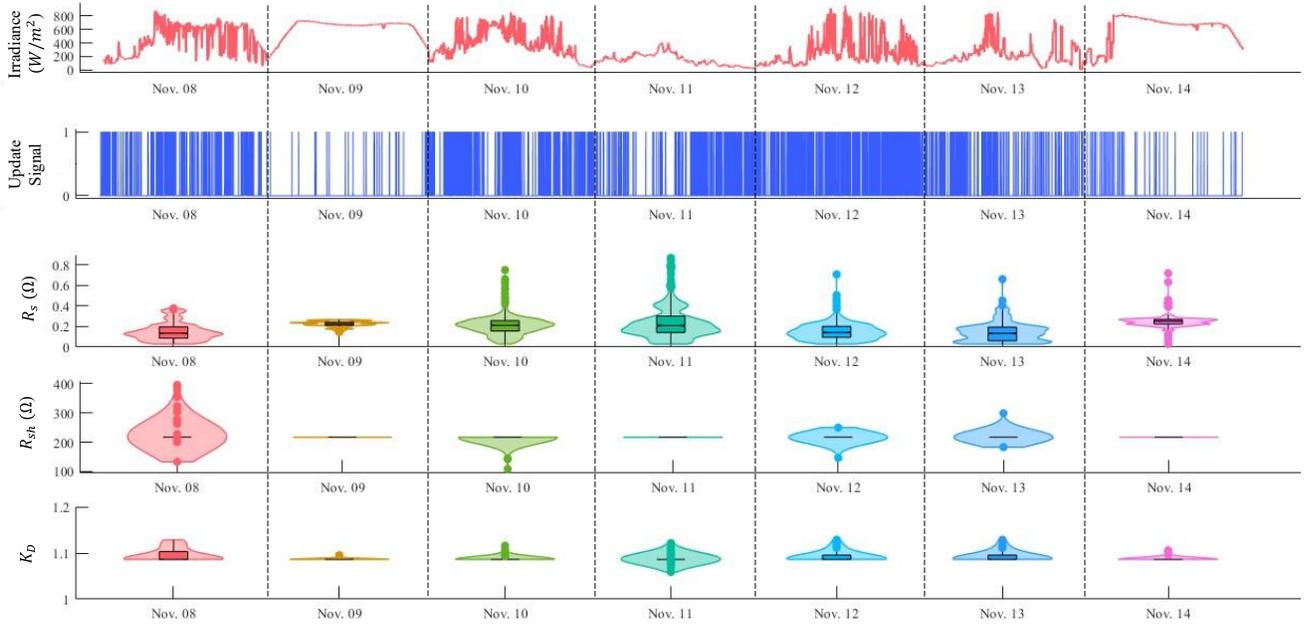

**Fig. 11.** Five model parameter's distribution for different types of days
\* No changes were observed in $I_{ph0}$ and $I_{s0}$, so they are omitted.

*C. PVDT Results under different weather conditions.*

In this section, we validate the performance of the proposed Event-trigger based Two-stage PSO for PV Farm data over a week, from November 8th to 14th, 2022. Fig. 11 presents the solar irradiance curve, the frequency of update signal occurrences, and the distribution of five model parameters during this period. On Nov. 9th, the irradiance curve gently slopes at the certain point of high value, indicating a clear sky and sunny day. The 14th experienced a brief cloudy period in the morning, followed by sunny conditions. On November 8th, 10th, 12th, and 13th, numerous ripples and spikes observed in the irradiance curve denote cloudy days with frequent partial shading moments. Moreover, November 11th was characterized by significantly low solar irradiance and can be defined as an overcast day with the sky fully covered by clouds.

A notable difference influenced by cloud cover is the number of update signal occurrences. The distribution of the five parameters varies with the cloudiness of the sky. On clear days like Nov. 9th, parameter fluctuations are notably minimal, evidenced by the low standard deviations, such as 0.028 ohms for $std(R_s)$ and 0.0013 for $std(K_D)$. In contrast, cloudy days exhibit a broader parameter distribution. This is likely due to the frequent partial shading events causing ripples and overshoots in the real-time measurement data, which serve as the basis for parameterization, thereby introducing significant variability in parameter estimation and increasing the number of updates. However, as demonstrated in Fig. 12, these fluctuations do not hinder achieving the desired PVDT performance.

As depicted in Table IV, the clearest day, Nov. 9th, required only 27 updates. In comparison, a typical day from 9 AM to 6 PM, updating every 10 seconds, would necessitate approximately 3,240 parameter updates, highlighting that event-trigger methods can reduce the update burden by more than a hundredfold while also achieving a lower current and voltage MAPE (0.22%, 0.20%) based on actual data.

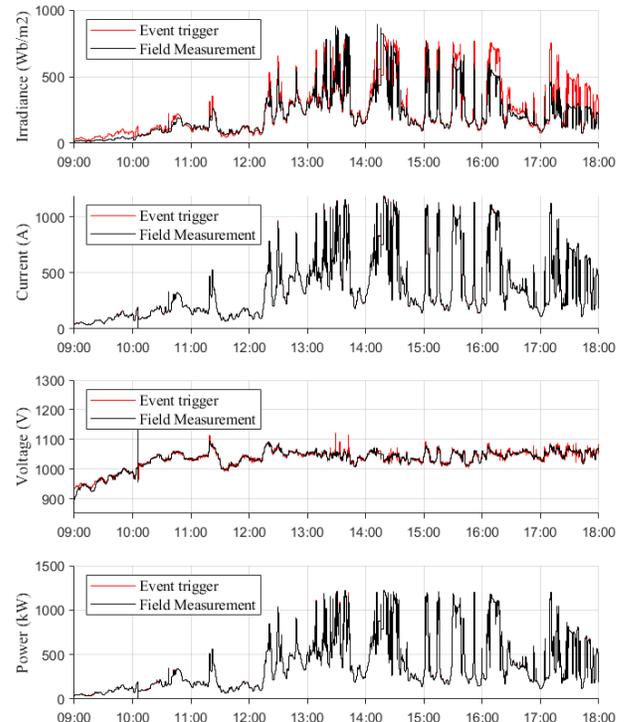

**Fig. 12.** PVDT result for a whole day – 2022.11.12

TABLE IV
PVDT RESULTS COMPARISON ACROSS WEATHER CONDITIONS

| Date | Weather Condition | Update Counts | $\Delta I$ (%) | $\Delta V$ (%) | $\Delta P$ (%) | $std(R_s)$ ($\Omega$) | $std(K_D)$ |
|---|---|---|---|---|---|---|---|
| 11.08 | Cloudy | 308 | 0.39 | 0.32 | 0.28 | 0.086 | 0.013 |
| **11.09** | **Sunny** | **27** | **0.22** | **0.20** | **0.10** | **0.028** | **0.0013** |
| 11.10 | Cloudy | 327 | 0.37 | 0.30 | 0.17 | 0.10 | 0.0026 |
| 11.11 | Overcast | 167 | 0.45 | 0.27 | 0.34 | 0.17 | 0.0047 |
| 11.12 | Cloudy | 430 | 0.51 | 0.40 | 0.26 | 0.079 | 0.011 |
| 11.13 | Cloudy | 152 | 0.40 | 0.28 | 0.24 | 0.10 | 0.011 |
| **11.14** | **Sunny** | **46** | **0.23** | **0.21** | **0.10** | **0.086** | **0.0015** |

Lastly, in Fig. 12, there are similarities and differences between the actual measured irradiance and the estimated equivalent irradiance. Initially, the measured power and the measured irradiance exhibit a moderately strong correlation until 3 p.m. (15:00). However, after 3 p.m., this correlation weakens significantly, with the measured power exceeds 1,000 $kW$ while the measured irradiance does not even reach 500 $W/m^2$, indicating a low correlation due to the shading effect and suboptimal tilt angle. In contrast, the estimated equivalent irradiance consistently maintains a strong correlation with the measured power throughout the day. This consistent correlation highlights the differences between estimated equivalent irradiance and measured irradiance. In this regard, using the measured irradiance directly, instead of the estimated equivalent irradiance, would significantly degrade the PVDT simulation performance.

## IV. CONCLUSION

The inaccuracies in second-by-second irradiance measurements lead to errors in the estimation of model parameters, adversely affecting the representation of actual large-scale PV systems. To cope with this matter, we introduce a novel PV modeling co-optimization method proposing a two-stage optimization approach to enhance the performance of real-time digital twins of PV farms. Our method not only demonstrates superior performance in various aspects by integrating and assessing different digital twin methodologies but also offers significant insights into its application in managing large-scale PV installations efficiently. Particularly, as shown in Fig. 12, our findings highlight that using measured irradiance directly can significantly degrade the simulation performance of PVDT, underscoring the effectiveness of our proposed method over traditional approaches.

Additionally, our proposed method utilizes an event-trigger approach, reducing communication exchange requirements without sacrificing parameterization performance. In future research, we will elucidate the changes in parameters due to failure and provide detailed analysis of PV's anomaly checks. Furthermore, a digital twin method for when curtailment occurs will also be proposed.